\documentclass[12pt]{article}

\usepackage{graphicx}
\usepackage{amsmath}
\usepackage{amssymb}

\newcommand{\eps}{\varepsilon}
\newcommand{\la}{\lambda}

\newcommand{\prt}{\partial}

\def\be{\begin {equation}}

\def\ee{\end {equation}}

\def\bea{\begin{eqnarray}}

\def\eea{\end{eqnarray}}

\begin{document}

\title{ \bf Asymptotic description of solitary wave trains in fully
nonlinear shallow-water theory}
\author{G.A.~El$^1$, \ R.H.J.~Grimshaw$^2$  \\
 Department of Mathematical Sciences, Loughborough University,\\
Loughborough LE11 3TU, UK \\
$^1$ e-mail: G.El@lboro.ac.uk \ \ \ $^2$ e-mail:
R.H.J.Grimshaw@lboro.ac.uk
 \\
 \\
 N.F.~Smyth\\
School of Mathematics and Maxwell Institute for Mathematical Sciences,\\
 University of Edinburgh,\\
The King's Buildings, Mayfield Road,\\
 Edinburgh, Scotland, EH9 3JZ, UK\\
e-mail:  N.Smyth@ed.ac.uk}

\date{}
\maketitle

\maketitle
\begin{abstract}
We derive an asymptotic formula for the amplitude distribution in a
fully nonlinear shallow-water solitary wave train which is formed as the
long-time outcome of the initial-value problem for the Su-Gardner
(or one-dimensional Green-Naghdi) system.  Our analysis is based on
the properties of the characteristics of the associated Whitham
modulation system which describes an intermediate ``undular bore'' stage of
the evolution.  The resulting formula represents a ``non-integrable''
analogue of the well-known semi-classical distribution
for the Korteweg-de Vries equation, which is usually obtained through the
inverse scattering transform. Our analytical results are shown to agree with
the results of direct numerical simulations of the Su-Gardner system.
Our analysis can be generalised to other weakly dispersive, fully nonlinear
systems which are not necessarily completely integrable.
\end{abstract}

\section{Introduction}
It is widely appreciated that, although completely integrable
systems may successfully capture many important features of
nonlinear dispersive wave propagation, they may fail to provide an
adequate description of large amplitude waves. Consequently,
significant efforts have recently been directed towards the derivation and
analysis of relatively simple models, enabling the quantitative
description of the propagation of fully nonlinear waves. Although
such ``intermediate'' models are typically not integrable using an
inverse scattering transform (IST), they often provide the possibility
of obtaining important particular solutions, as well as having other
advantages compared to the full physical system from the
viewpoint of numerical simulations.

One such system, the so-called Green-Naghdi (GN) equations,
describes large-amplitude shallow water waves \cite{gn76}
and also appears in a number of other fluid dynamics contexts,
such as ``continua with memory'' \cite{gav94}, solar
magnetohydrodynamics \cite{dellar03}, bubbly fluid flows \cite{gt01}
and the dynamics of short capillary-gravity waves \cite{manna2}.
The two-layer versions of the GN system \cite{cc99, ost, bgt07}
provide a broad field for the modelling of large amplitude
interfacial waves. It should be noted that the one-dimensional version
of the GN system, which is the main subject of this paper, was
originally derived by Su and Gardner \cite{sg69} using a long-wave
asymptotic expansion of the full Euler equations for irrotational
flow (see also El, Grimshaw \& Smyth \cite{egs06}), while the original 2D
GN system was derived using the ``directed fluid sheets'' theory,
which does not require a formal asymptotic expansion, but instead
imposes the condition that the vertical velocity has only a linear
dependence on the vertical $(z)$ coordinate, and simultaneously
assumes that the horizontal velocity is independent of $(z)$. In
this paper we shall use the term SG system which appears historically
more correct, at least for one-dimensional dynamics.   The SG system has the
form
\begin{eqnarray}
&&\eta_t+(\eta u)_x = 0\, ,  \label{gn}\\
&&
u_t+uu_x+\eta_x=\frac{1}{\eta}\left[\frac{1}{3}\eta^3(u_{xt}+uu_{xx}-
(u_x)^2)\right]_x \, , \nonumber
\end{eqnarray}
where, in the context of shallow-water waves, $\eta$ is the total
depth and $u$ is the layer-mean horizontal velocity; all variables
are non-dimensionalised by their typical values. The first equation
is the exact equation for conservation of mass and the second
equation can be regarded as an approximation to the equation for
conservation of horizontal momentum.  The system (\ref{gn}) has the
typical structure of the well-known Boussinesq-type systems for
shallow water waves, but differs from them in retaining full
nonlinearity in the leading-order dispersive term on the
right-hand side of (\ref{gn}).  We stress that
there is no limitation on the amplitude assumed in the derivation of
(\ref{gn}). This, along with the appearance of this system in
different physical contexts, suggests that equations (\ref{gn})
represent an important mathematical model for understanding general
properties of fully nonlinear fluid flows beyond the strict shallow-water
limit. For this reason, it is instructive to study its
solutions for the full range of amplitudes, although in the
particular context of shallow-water waves the system (\ref{gn}), as
for any layer-mean model, is unable to reproduce the effects of wave
overturning and becomes nonphysical for amplitudes greater than some
critical value.

The system (\ref{gn}) has a solitary wave solution of the form
\begin{equation} \label{sol}
\eta=\eta_0+a \ \hbox{sech}^2\left(\frac{ \sqrt{3a}}{\eta_0\sqrt
{\eta_0+a}}(x-ct) \right) \, ,
\end{equation}
where the solitary wave speed $c$ is connected to the amplitude $a$ by the
relationship
\begin{equation} \label{cs}
 c= u_0 + \sqrt{\eta_0 +a} \, .
\end{equation}
Here $u_0$ and $\eta_0$ are the background flow parameters. We note
that formula (\ref{cs}) appears in Rayleigh \cite{ray}.

The asymptotic reduction of the SG system (\ref{gn}) for weakly
nonlinear waves is obtained by using the standard scaling
$\eta=\eta_0 + \delta \zeta $, $u=u_0 + \delta U$, $x -c_0 t \to
\delta^{-1/2}x$, $t \to \delta^{-3/2} t$, where $\delta$ is a
small parameter and $c_0 = u_0 + \sqrt{\zeta_0 }$ is the linear longwave
speed.  For uni-directional propagation, this leads to the KdV
equation (see, for instance, Johnson \cite{johnson02})
\begin{equation}\label{kdv}
 \zeta_t  +\frac{3}{2} \zeta \zeta_x+\frac{1}{6}\zeta_{xxx} = 0\,.
\end{equation}
With the further scaling
$x^{\prime }  = \sqrt{6}x \,,  t^{\prime }=\sqrt{6}t \,,  v= 3\zeta /2$,
(\ref{kdv}) reduces to the standard KdV form
\begin{equation}\label{kdv1}
v_t + vv_x+v_{xxx}=0\, ,
\end{equation}
on omitting the primes.   Then if one considers sufficiently rapidly
decaying positive initial data for (\ref{kdv1})
\begin{equation}\label{ic1}
\qquad v(x,0)=v_0(x) \ge 0 \, , \qquad v_0 \to 0 \ \ \hbox{as} \ \
|x| \to \infty\, ,
\end{equation}
where $v_0(x)$ is a continuous ``one-hump'' function, then
as $t \to \infty$ the solution asymptotically consists of a finite number of
solitons plus dispersive radiation.

The associated linear spectral problem for (\ref{kdv1}) is given by
the Schr\"odinger equation
\begin{equation}\label{spectral}
6 \psi_{xx} + (v(x,t) - \la )\psi=0 \,.
\end{equation}
Note that this is in non-standard form due to the factor $6$ in the first term.
Then the parameters of the solitons and the radiation are found from the IST
using the initial condition $v_{0}(x)$ (\ref{ic1}) for $v$ in (\ref{spectral}).
In particular, a soliton with amplitude $a_n$ corresponds to the
eigenvalue $\la_n = a_n/2$ of the Schr\"odinger operator.

If one considers a {\it large-scale} initial distribution such that
$\epsilon = 1/( A^{1/2}w) \ll 1$, where $A \propto \max v_0$, and
$w$ is the typical width of $v_0(x)$, then the contribution of the
radiation is exponentially small in $\epsilon$ and as $t \to
\infty$ the asymptotic outcome consists only of a large ($N \propto
\epsilon ^{-1}$) number of solitons, whose amplitudes can be found
from the Bohr-Sommerfeld semi-classical quantization rule,
\begin{equation}\label{BS}
\frac{1}{\sqrt{6}}\int_{x_1}^{x_2} \sqrt{v_{0}(x) - \la } \ dx
=\pi(n-\frac{1}{2})\, , \quad n=1,2,3, \dots, N \, .
\end{equation}
Here the integral is taken between the turning points $x_{1,2}$
defined as the roots of the equation $v_0(x)=\lambda$. For every
given $n$ formula (\ref{BS}) yields a bound state $\lambda_n > 0$; \,
$A\approx \lambda_1>\lambda_2> \dots
>\lambda_N \approx 0$ with $N$ the total number of bound states in the
potential $v_0(x)$. In the KdV context the distribution (\ref{BS}),
through $a_n=2\lambda_n$, gives the amplitude of the $n$-th soliton in
the soliton train as $t \to \infty$. For the largest amplitude soliton we have
the classical result
\begin{equation}\label{amax}
a_{\max}=2\lambda_1=2A \, .
\end{equation}
On the other hand, for a given $\lambda \in [0, A]$, the formula
(\ref{BS}) determines the total number of bound states $n$ in the
spectral interval $(\lambda, A)$ as
\begin{equation}\label{}
n= \hbox{i.p.} \ (F(\lambda)+\frac12)\, , \qquad F(\lambda) =
\frac{1}{\sqrt{6} \pi}\int_{x_1}^{x_2} \sqrt{v_0(x)- \lambda} \ dx \, .
\end{equation}
Here ``$\hbox{i.p.} \ (\cdot)$'' denotes the integer part.

For large $n$ the bound states are located close to each other and
we can introduce a continuous amplitude function $a=2\lambda$. Then
differentiating (\ref{BS}) with respect to $a$, one obtains the
distribution function for soliton amplitudes in the train
\begin{equation}\label{karp}
f(a) = \left |\frac{dn}{da} \right |=\frac{1}{4\pi}
\frac{1}{\sqrt{6}}\int_{x_1}^{x_2} \frac{dx}{\sqrt{v_0(x)- a/2}}\, ,
\end{equation}
so that $f(a)da$ yields the number of solitons having
amplitudes in the interval $(a,a+da)$. The total number of solitons
in the soliton train is estimated by setting $\lambda=0$ in
(\ref{BS}) (alternatively, one can integrate $f(a)$ from $0$ to
$a_{\max}$)
\begin{equation}\label{N1}
N \cong \frac{1}{\sqrt{6} \pi} \int_{-\infty}^{+\infty}
\sqrt{v_0(x)} \ dx  \sim A^{1/2}w\, .
\end{equation}
The distribution (\ref{karp}) was originally introduced into soliton
theory by Karpman \cite{karp67} (see also Whitham \cite{wh74} and Karpman
\cite{karp75}).  More recently, distributions of this type have been obtained,
also from the associated linear spectral problem, for the defocusing
nonlinear Schr\"odinger equation \cite{kku02} and for the Kaup-Boussinesq
system \cite{kku03}.

One can see that integrability of the KdV equation plays a crucial
role in the derivation of the formulae (\ref{BS})--(\ref{N1}) through the
eigenvalue problem for the associated linear Schr\"odinger equation
(\ref{spectral}).  At the same time one can observe that the formulae
obtained correspond to the semi-classical approximation and that this very
approximation can be applied directly to the KdV equation, by-passing
the IST formalism. It is well-known (see Lax and Levermore \cite{llv94})
that the semi-classical limit of the KdV equation with
decaying initial data leads to the Whitham modulation system, which
can be derived by a direct averaging of the KdV conservation laws
over nonlinear wavepackets \cite{wh65} or from a multiple-scale
perturbation analysis \cite{luke,dm82}.
Therefore one could expect that the results (\ref{amax})--(\ref{N1})
could be obtained within the framework of the modulation
theory alone. Indeed, it was shown in Gurevich {\em et al} \cite{gur92}
and El \& Grimshaw \cite{eg02} that the
long-time asymptotics of exact solutions to the KdV-Whitham
equations with decaying initial data agrees, to first order in
$t^{-1}$, with Karpman's formula (\ref{karp}). Still, modulation
solutions in the cited papers rely on the integrability of the KdV
equation as they employ the presence of Riemann invariants for the
associated Whitham system \cite{wh65}, and this latter property
is due to the unique finite-gap spectral structure of periodic (or,
more-generally, quasi-periodic) solutions of the KdV equation
\cite{ffm}.

An important theme in the original work of Whitham \cite{wh65} is that,
unlike the IST, the nonlinear modulation approach can be applied to
nonintegrable dispersive systems provided a minimal structure is
present, that is the existence of periodic solutions characterised by a
certain number of parameters, and the availability of a limited number of
conservation laws. The resulting modulation system consists of
hydrodynamic-type equations and is hyperbolic in many cases, which
implies that it can be treated using classical methods of
characteristics theory (see, for instance, Courant \& Hilbert
\cite{Cour}). This opens the possibility of obtaining some analytical results
for non-integrable dispersive wave systems in the framework of modulation
theory.

Indeed, it was recently shown in El \cite{el05} that the availability of
a number of important exact results for the single-phase modulation
theory is, in fact, due to certain very general properties of
modulation systems and is not connected with the presence of the
Riemann invariant structure. These properties have been used in El,
Grimshaw \& Smyth \cite{egs06} to derive the main parameters of
large-amplitude shallow-water undular bores evolving from an initial
step for the SG system (\ref{gn}). In the present paper, we extend
the modulation analysis of El \cite{el05} and El, Grimshaw \& Smyth
\cite{egs06} to obtain an analogue of Karpman's formula (\ref{karp}) and its
consequences (\ref{amax}) and (\ref{N1}) for the SG system for
an initial disturbance in depth and velocity that decays at infinity.
Our approach is based on an assumption, confirmed by numerical
solutions, that the main qualitative features of the KdV evolution
of large-scale disturbances, such as the formation of a single-phase
undular bore and its further evolution into a solitary wave train
with a negligibly small contribution of the linear radiation, remain
present in the SG model. Our analytical results will be supported by
comparison with full numerical solutions of the SG system.

\section{Asymptotic formula for the KdV equation derived from
modulation theory}

\subsection{Characteristic integrals of the modulation system}

It is instructive to start with a demonstration of our general
approach to an asymptotic description of solitary wave trains by
using the KdV equation as a ``test'' example. Our ultimate goal here
is to derive Karpman's formula (\ref{karp}) without invoking the
integrability properties of the KdV equation and its modulation
system. This derivation will then serve as a prototype for
calculations for the SG system, which is genuinely non-integrable.

We consider the initial-value problem (\ref{kdv1}) and (\ref{ic1}).
Without being too restrictive, we also assume that $v_0(x)=0$
for $x \ge 0$, which will simplify the following analysis. As before, we
consider a large-scale initial disturbance, so that $(A^{1/2}w)^{-1} \ll 1$,
where $A=\max{(v_0)}$ and $w$ is the characteristic width of the
initial hump.

The wave evolution from (\ref{kdv1}) and (\ref{ic1}) leads to wave breaking
at some $t= t_b$, which can be taken to be $t_{b}=0$ without loss of
generality, which is then resolved by an undular bore (which is also often
called a dispersive shock wave), which is an expanding nonlinear modulated
wave train with a distinctive spatial structure.  Near the leading edge of
the undular bore the oscillations appear to be close to successive
solitary waves, while in the vicinity of the trailing edge they are
nearly linear.  An asymptotic similarity modulation solution for the
undular bore evolving from an initial distribution in the form of a
sharp step was first obtained by Gurevich and Pitaevskii (GP) \cite{GP}
using the Whitham method of averaging over periodic nonlinear
wavetrains \cite{wh74,wh65} and then rigorously recovered in the
framework of the semiclassical IST formalism by Lax, Levermore and
Venakides (see their review \cite{llv94} and references therein).  In
contrast to the Lax-Levermore-Venakides global construction, the
direct modulation approach of GP does not rely on the IST and thus
has the advantage of potential applicability to non-integrable
systems.  This advantage was realised in El \cite{el05} for the
simplest class of problems with step-like initial conditions where
the modulation solution for the undular bore represents an expansion fan,
similar to the GP solution.

For the case of positive initial data, which decay at infinity, the
whole initial disturbance decomposes, at very large times $t \gg
\eps^{-1}$, into a chain of solitons with a certain amplitude
distribution.  As was mentioned in the Introduction, the semi-classical
IST approach leads to Karpman's formula (\ref{karp}) for this
distribution.  However, it transpired that the same result can be
obtained using the derivation of the intermediate modulation solution
for the undular bore and then considering its long-time asymptotic
behaviour \cite{gur92,eg02}. In this
connection it should be mentioned that the modulation dynamics in
these problems with decaying initial data is not self-similar, so the
original Gurevich-Pitaevskii \cite{GP} theory required significant
development in order to be applied to such problems.  It transpires that
further development is required to extend modulation analysis to
non-integrable systems with decaying initial data. Below we present
the GP formulation for the KdV equation in a form convenient for
further generalistion to non-integrable systems.

The local waveform of the undular bore is described by the
single-phase periodic solution of the KdV equation travelling with
constant velocity $c$, that is $v(x,t)=v(\theta)$, $\theta=x-ct$. This
solution is specified by the ordinary differential equation
\begin{equation}\label{tr}
(v_\theta)^2= Q(v) \, , \quad  Q(v)=
\frac{1}{3}(v-v_1)(v-v_2)(v_3-v)\, , \quad v(\theta +
2\pi/k)=v(\theta)\, ,
\end{equation}
$v_3 \ge v_2 \ge v_1$ being constants of integration. The phase
velocity $c$, the amplitude $a$, the wavenumber $k$ and the mean
$\overline v$ are expressed in terms of the polynomial roots $v_j$
as
\begin{equation}\label{L}
c=\frac{1}{3}(v_1+v_2+v_3), \quad a=v_3-v_2, \quad \quad k={\pi}
\left(\int\limits^{v_3}_{v_2}\frac{dv}{\sqrt{Q(v)}}\right)^{-1},
\quad \overline v = \frac{\pi}{k}\int\limits^{v_3}_{v_2}\frac{v
dv}{\sqrt{Q(v)}},
\end{equation}
while the wave frequency is $\omega(\overline v, k, a)= kc$.
When $a \to 0$ the solution of (\ref{tr}) turns into
small amplitude harmonic waves with the dispersion relation
\begin{equation}\label{dr0}
\omega_0(\overline v, k)=\omega( \overline v, k, 0) = k \overline
v-k^3 \, .
\end{equation}
In an opposite extreme, when $k \to 0$, the travelling wave
$u(\theta)$ transforms into a soliton with its velocity depending on
the amplitude as
\begin{equation}\label{ca}
c_{sol}= \overline v +\frac{a}{3}
\end{equation}
If one allows slow dependence of $v_1$, $v_2$ and $v_3$ or, equivalently,
$\overline v$, $k$, and $a$ on $x$ and $t$, one arrives at the modulation
system which can be derived by averaging two of the KdV conservation
laws over the periodic wave family (\ref{tr})
\begin{equation}\label{wh2}
\frac{\partial \overline v}{\partial t}+  \frac{\partial }{\partial
x}\left(\frac{\overline{v^2}}{2}\right)=0 \, , \quad
\frac{\partial}{\partial t}\left( \frac{\overline{v^2}}{2}\right)+
\frac{\partial }{\partial x}\left(\overline {\frac{v^3}{3}
-\frac{v_\theta^2}{2} + vv_{\theta \theta}} \right)=0 \,
\end{equation}
and then closing (\ref{wh2}) with the equation for ``conservation of
waves'' (see Whitham \cite{wh74})
\begin{equation}\label{wc}
\frac{\partial k}{\partial t} + \frac{\partial \omega (\overline v,
k, a)}{\partial x}=0\, .
\end{equation}
We note that the detailed expressions for the averages in
(\ref{wh2}) will not be needed. To describe the evolution of
modulations in an undular bore the system (\ref{wh2}) and (\ref{wc})
must be equipped with matching conditions ensuring continuity of
the mean $\overline u$ at the undular bore edges $x^{\pm}(t)$
\cite{GP}
\begin{equation}
\begin{array}{l}
x=x^-(t):\qquad a=0\, , \ \ \overline v = \beta^-(t)\, , \\
x=x^+(t):\qquad k=0\, , \ \ \overline v = \beta^+(t) \, .
\end{array}
\label{gp}
\end{equation}
Here $\beta^{\pm}(t) = \beta(x^{\pm}(t), t)$, where $\beta(x,t)$ is
the solution of the Hopf equation
\begin{equation}\label{hopf}
\beta_t+\beta\beta_x=0\, , \quad
\beta(x,0)=v_0(x)\,  ,
\end{equation}
which represents the dispersionless limit of the KdV equation
(\ref{kdv}) and is valid outside the oscillatory region. As we have
assumed that $v_0(x ) = 0$ for $x \ge 0$, we have $\beta^+ = 0$.

The trailing $x^{-}(t)$ and the leading $x^{+}(t)$ edges of the
undular bore represent free boundaries defined by the kinematic
conditions
\begin{equation}\label{kin}
\frac{dx^-}{dt}= c_g(\beta^-, k^-) \,   , \qquad
\frac{dx^+}{dt}=c_{sol}(\beta^+, a^+)\, ,
\end{equation}
where $c_g=\partial \omega_0 / \partial k$ is the group velocity of
linear waves and $k^-=k^-(t)=k(x^-(t),t)$, $a^+=a^+(t)=a(x^{+}(t),
t)$. It is important to note that, according to general properties
of quasi-linear hyperbolic systems (see, for instance, Courant \&
Hilbert \cite{Cour}), the curves $x^{\pm}(t)$, being the lines which match
two analytically different solutions, must coincide with
characteristics of the modulation system (\ref{wh2}) and (\ref{wc}).
Indeed, as we shall see, the kinematic conditions (\ref{kin}) and
the requirement that the undular bore boundaries must be
characteristics of the modulation system are equivalent.

We now consider the modulation equations in two distinguished
limits: as $a \to 0$ and as $k \to 0$--- corresponding to the wave
regimes at the trailing and the leading edges of the undular bore
respectively. When $a \to 0$ the oscillations do not contribute to
the averaging, so $\overline {F(v)} = F(\overline v)$, where $F(v)$
is an arbitrary function, and the modulation system must reduce to
(see El \cite{el05} for details)
\begin{equation}\label{red1}
a=0\, , \qquad \overline v _t + \overline v  \cdot \overline v _x =0
\, , \qquad k_t + (\omega_0(\overline v, k))_x = 0 \, .
\end{equation}
The system (\ref{red1}) has two families of characteristics,
$dx/dt=\overline v$ and $dx/dt=\partial \omega_0/\partial k$. The
first family is consistent with the characteristics of the
``external'' Hopf equation (\ref{hopf}) which transfers initial data
from the $x$ axis into the undular bore region in the
$(x,t)$-plane, while one of the characteristics of the second family
specifies the trailing edge of the undular bore (see the first
kinematic condition (\ref{kin})). Then, according to the general
properties for the prescription of Cauchy data on characteristics
(see, for instance, Whitham \cite{wh74}), one cannot specify the values of
$\overline v$ and $k$ at the trailing edge independently.  Of course,
a similar statement is true for the leading edge as well, and this
is why the matching conditions (\ref{gp}) are sufficient to
determine the evolution of an undular bore regardless of the fact
that they involve less variables than the modulation system
(\ref{wh2}). The admissible combinations of the values of
$\overline v$ and $k$ on a characteristic with $a=0$ are found by a
substitution of $k= k(\overline v)$ into (\ref{red1}), which leads to
the ordinary differential equation
\begin{equation}\label{ode1}
\qquad \frac{dk}{d\overline v}=\frac{\partial \omega_0 / \partial
\overline v}{\overline v - \partial \omega_0 / \partial k}.
\end{equation}
Substituting the linear dispersion relation (\ref{dr0}) into
(\ref{ode1}) one readily obtains the characteristic integral
\begin{equation}
\label{ku}
k(\overline v)= \sqrt{\frac{2}{3}
(\overline v - \lambda_1)}  \, ,
\end{equation}
where $\lambda_1$ is an arbitrary constant.

We now consider the soliton limit $k \to 0$.  Here the wavelength $2\pi/k$
tends to infinity, so that the contribution of oscillations to the mean value
$\overline v$ vanishes, and, similarly to the case of vanishing
amplitude, we have $\overline{F(v)}=F(\overline{v})$. Hence, we
arrive, again, at the Hopf equation for $\overline v$,
\begin{equation}\label{hopf1}
k=0: \qquad  \overline v _t + \overline v \cdot \overline v _x =0\,.
\end{equation}
We finally pass to the limit as $k \to 0$ in the wave
conservation law (\ref{wc}).  This limiting transition, unlike that
as $a \to 0$, is a singular one, so that it requires a more careful
treatment. Firstly we note that the wave conservation law is
satisfied identically for $k=0$, so we need to take into account
higher order terms in the expansion of (\ref{wc}) for small $k$. It
is then convenient to introduce a ``conjugate wave number'' (cf.\
(\ref{L}))
\begin{equation}\label{conjk}
    \widetilde{k}=\pi\left(\int_{v_1}^{v_2}
    \frac{d v}{\sqrt{Q(v)}}\right)^{-1}
\end{equation}
instead of the amplitude $a$ and the ratio
$\Lambda=k/\widetilde{k}$ instead of the original wave number $k$,
so that the parameters $(\overline{v}, \Lambda, \widetilde{k})$
form a new set of modulation variables which is more convenient for
the consideration of the vicinity of the soliton edge of an undular
bore than our original set $(\overline{v}, k, a)$. The variable
$\widetilde{k}$ can be considered as the wavenumber of a ``conjugate
travelling wave'' specified by the equation
\begin{equation}\label{conjtrav}
    \left(\frac{d v}{d\widetilde{\theta}}\right)^2=-Q(v)\, , \qquad
v(\widetilde \theta + 2\pi/\widetilde k)= v(\widetilde \theta),
\quad u \in [v_1, v_3]\, ,
\end{equation}
where $Q(v)$ is the same as in Eq.~(\ref{tr}) and $\widetilde \theta
= \tilde x - c \tilde t$ is a new phase variable characterised
by the same phase velocity $c$. Equation (\ref{conjtrav}) specifies
periodic solutions of the ``conjugate'' KdV equation
\begin{equation}\label{kdv2}
v_{\tilde t} + v v_{\tilde x} - v_{\tilde x \tilde x \tilde x}=0 \,,
\end{equation}
which can be obtained from (\ref{kdv1}) by the change of variables
$x \mapsto i \tilde x$ and $t \mapsto i \tilde t$. It is not
difficult to infer from (\ref{conjk}) and (\ref{conjtrav}) that in
the limit $v_1 \to v_2$ (i.e.\ $k \to 0$) $\widetilde k$ has the
meaning of an inverse soliton width (or ``soliton wavenumber'')
defined by the asymptotic behaviour in the soliton tails $v -
\overline v \sim \exp (- \widetilde k |\theta|)$ as $ |\theta| \gg 1
$. Moreover, it follows from (\ref{conjtrav}) that the dependence of
the soliton speed, which coincides with the value of the phase
velocity $c$ evaluated in the limit as $v_1 \to v_2$ on its inverse
width (and, therefore, from (\ref{ca}), on the amplitude $a$),
follows from the conjugate linear dispersion relation $\widetilde
\omega = \widetilde \omega_0(\widetilde k)$. The latter is obtained
from the linear dispersion relation (\ref{dr0}) by the change of
variables $k \mapsto i \widetilde k$ and $\omega_0 \mapsto -i \tilde
\omega_0$, i.e.\ $i\widetilde{\omega}_0(\widetilde
k)=\omega_0(i\widetilde k)$.  We thus have
\begin{equation}\label{conjomega}
\widetilde \omega_0(\widetilde k) = \widetilde k \overline v + \widetilde k^3
\end{equation}
and, comparing with (\ref{ca}), we obtain the well known relationship
between the KdV soliton amplitude and its inverse width
\begin{equation}\label{ak}
k=0: \qquad \frac{a}{3}= \widetilde k^2\, .
\end{equation}
We are now ready to study the asymptotic expansion of the wave
conservation law for $k \ll 1$. First we substitute $k=\Lambda
\widetilde{k}$ into Eq.~(\ref{wc}) to obtain the equivalent
representation
\begin{equation}\label{wc1}
\tilde k  \Lambda_z + \tilde \omega \Lambda_x+ \Lambda ( \tilde k_z
+ \tilde \omega_x )=0 \, ,
\end{equation}
where $\widetilde \omega = c \widetilde k$. Next we consider
Eq.~(\ref{wc1}) for small $\Lambda\ll 1$ and assume that $\Lambda
\ll \Lambda_t, \Lambda_x$ for the solutions of interest.  We note that this
is known to be the case for modulation solutions describing
undular bores in weakly dispersive systems, where at the soliton
edge one has $k \to 0$, but $|k_x|, |k_z| \to \infty$ (see El \cite{el05}
for a general discussion of this behaviour and Gurevich and
Pitaevskii \cite{GP} for the detailed calculations in the KdV case).
Then to leading order we obtain the characteristic equation
\begin{equation}\label{cl}
\frac{\partial \Lambda}{\partial t} + \frac{\widetilde
\omega_0}{\widetilde k} \frac{\partial \Lambda}{\partial x}=0 \, ,
\end{equation}
or
\begin{equation}\label{c2}
\Lambda= \Lambda_0  \qquad \hbox{on} \qquad
\frac{dx}{dt}=\dfrac{\widetilde \omega_0 (\overline v, \widetilde
k)}{\widetilde k}\, ,
\end{equation}
where $\Lambda_0 \ll 1$ is a constant. In particular, when
$\Lambda_0=0$ (that is $k =0$), the characteristic (\ref{c2})
specifies the leading edge of the undular bore (cf.\ (\ref{kin})).
Now, considering a restriction of equation (\ref{wc1}) to the
characteristic family $dx/dz=\widetilde \omega_0/\widetilde k$ and
using that $\widetilde \omega = \widetilde \omega_0$ to leading
order, we obtain
\begin{equation}\label{cws}
k=0: \qquad \widetilde k_t+(\widetilde \omega_0)_x=0 \qquad
\hbox{on} \qquad \frac{dx}{dt}=\dfrac{\widetilde \omega_0 (\overline
v, \widetilde k)}{\widetilde k}\, .
\end{equation}
We note that the equation $\widetilde k_t+(\widetilde \omega_0)_x=0$
arises as a ``soliton wavenumber'' conservation law in the
traditional perturbation theory for a single soliton (see, for
instance, Grimshaw \cite{g79}), but to be consistent with full modulation
theory it should be considered along the soliton path
$dx/dt=c_{sol}= \widetilde \omega_0/\widetilde k$.

Since $\overline v$ and $\widetilde k$ cannot be specified
independently on a characteristic, there should exist a local
relationship $\widetilde k(\overline v)$ consistent with the system
(\ref{hopf}) and (\ref{cws}). Substituting $\widetilde k=\widetilde
k(\overline v)$ into (\ref{cws}) and using (\ref{hopf}) we obtain
an equation for $\widetilde k$ similar to equation (\ref{ode1}) for
$k$ obtained earlier in the opposite limit as $a \to 0$
\begin{equation}\label{ode2}
    \frac{d\widetilde k}{d \overline v}=\frac{\prt\widetilde{\omega}_0/
\prt\overline v}
    {\overline v -\prt\widetilde{\omega}_0/\prt \widetilde k} .
\end{equation}
Substituting $\widetilde \omega_0$ (\ref{conjomega}) into
(\ref{ode2}) one readily integrates to obtain
 \begin{equation}\label{kappau}
\widetilde k(\overline v)= \sqrt{\frac{2}{3}(\lambda_2 - \overline
v)} \, ,
\end{equation}
where $\lambda_2$ is an arbitrary constant.

Now we use the fact that the linear wave packet at the trailing edge
and the lead solitary wave are not independent, but rather are
constrained by the condition of being parts of the same undular
bore. So, if one considers a pair of integrals (\ref{ku}) and
(\ref{kappau}) with certain constants $\lambda_1$ and $\lambda_2$ in
the context of the {\it same} modulation solution, then $\lambda_1$
and $\lambda_2$ cannot be set independently. One can see from
(\ref{ku}) that if $\overline v = \lambda_1$, then $k=0$, which then
must be consistent with equation (\ref{kappau}) derived for the
soliton configuration. But, at the same time, equation (\ref{ku})
corresponds to $a=0$, which, together with $k=0$, implies
$\widetilde k=0$ (see (\ref{ak})). So setting $\overline v =
\lambda_1$ in (\ref{kappau}) immediately yields $\lambda_1 =
\lambda_2 \equiv \lambda$ and we therefore arrive at the set of two
consistent characteristic integrals of the modulation system
\begin{equation}\label{int1}
I_1=\left \{a=0, \ \ k=\sqrt{\frac{2}{3}(\overline v -
\lambda)}\right \} \quad \hbox{on} \quad \frac{dx}{dt} =
\frac{\partial \omega_0}{\partial k},
\end{equation}
\begin{equation}\label{int2}
 I_2=\left \{ k=0, \ \ \widetilde k
= \sqrt{\frac{2}{3}(\lambda - \overline v)}\right \} \quad \hbox{on}
\quad \frac{dx}{dt}=\frac{\widetilde \omega_0 (\overline v,
\widetilde k)}{\widetilde k}\, .
\end{equation}
One should add that both $k$ and $\widetilde k$ are required to be
real, so the integrals $I_{1,2}$ involve different parts of the
domain of the function $\overline v(x,t)$.

\subsection{Total number of solitons}

We start with the determination of the total number of solitons
generated by the decay of the given initial disturbance (\ref{kdv1})
and then proceed by obtaining a more detailed description of a
soliton train by finding the distribution function for the soliton
amplitudes in terms of the initial data.

We consider the wave conservation law (\ref{wc}).  For the case of
the decaying initial profile (\ref{kdv1}) one obviously has $\omega \to
0$ as $|x| \to \infty$ for all $t$, and hence equation (\ref{wc}) implies
conservation of the total number of oscillations (wave crests)
\begin{equation}\label{N2}
N \cong \frac{1}{2\pi}\int^{+\infty}_{-\infty} k dx =
\hbox{constant}\, .
\end{equation}
We use an approximate equality sign here due to the inherent
asymptotic character of modulation theory.  Next we note that,
qualitatively, the process of soliton generation during the
evolution of the large-scale, decaying initial profile (\ref{ic1}) can
be described as follows (see Gurevich, Krylov \& Mazur \cite{gur89} for
a quantitative justification): each wave crest is generated at the
trailing edge as a vanishingly small amplitude linear wave and
asymptotically, as $t \to \infty$, transforms into a soliton.
Formula (\ref{N2}) then can be used for the evaluation of the total
number of solitons in the eventual soliton train.

To evaluate the integral in (\ref{N2}) one needs to know the function
$k(x,t)$ for all $x$ at any particular $t$, say at $t=0$.
The difficulty with the traditional Gurevich-Pitaevskii approach to the
undular bore description is that the wave number $k$ in this approach is
defined only within the undular bore region $[x^-, x^+]$. However, one can
extend the notion of the wavenumber to the entire $x$-axis by
defining the function $k(x,t)$ in such a way that its behaviour
outside the undular bore region is consistent with the
prescribed values of $k$ along the leading and trailing edges
$x^{\pm}(t)$ for all $t$.

At the leading edge $x^+(t)$ we have $k=0$ for all $t$.  Therefore
we define $k = 0$ for $x \ge x^+(t)$. Hence here we have
\begin{equation}\label{k01}
k(x,0)=0 \qquad \hbox{for}  \qquad x \ge 0.
\end{equation}
At the trailing edge $x^-(t)$ the amplitude of the wave $a=0$, and
since the trailing edge represents a characteristic of the
modulation system, the value of the wavenumber $k^-=k(x^-,t)$ is
determined by the boundary value of $\overline v (x^-, t) =
\beta^-$ from the characteristic integral $I_1$ in (\ref{int1})
for a certain $\lambda$ (see (\ref{gp}) and (\ref{hopf}) for the
definitions of $\beta^{\pm}(t)$). To be consistent with the matching
conditions (\ref{gp}), we need to set $\lambda =0$.  Indeed since at
the leading edge $\overline v (x^+, t)= \beta^+=0$, then the
dependence $k^-(\beta^-)$ must correctly reproduce the value $k=0$
for the case when $\beta^-=\beta^+=0$, implying $a=0$ and $k=0$
simultaneously, which by (\ref{ak}) also implies $\widetilde k=0$.
So we obtain
\begin{equation}\label{k-}
k^-=\sqrt{(2/3)\beta^-}.
\end{equation}

Now upstream of the undular bore, that is $x \le x^-$, the function
$\beta(x,t)$ satisfies the Hopf equation (\ref{hopf}) with the
initial condition $\beta(x,0)=v_0(x)$, i.e. implicitly
$\beta(x,t)=v_0(x- \beta t)$. Therefore for $x \le x^-(t)$ we need
to define the wavenumber as $k(x,t) = k^-(\beta(x,t))$, where
$\beta(x,t)$ is the aforementioned simple-wave solution. This
extension basically implies that we assume that upstream of the
undular bore, where $a=0$, the relationship (\ref{int1}) holds {\it
everywhere} (not only for a special family of characteristics
$dx/dt=\partial \omega_0\partial k$). Indeed it can be readily seen
that $k=\sqrt{(2/3)\overline v}$ is a solution of the reduced
modulation system (\ref{red1}) regardless of the particular
characteristic family. Thus at $t=0$ we obtain
\begin{equation}\label{k02}
k(x,0)=k^-(\beta(x,0))=\sqrt {\frac{2}{3}v_0(x)}\, \qquad \hbox{for}
\qquad x \le 0 \, .
\end{equation}
We note that since $v_0(x \ge 0)=0$ this formula is also
consistent with our definition (\ref{k01}) for $k(x,0)$ for $x\ge
0$, so that (\ref{k02}) gives the function $k(x,0)\equiv k_0(x)=\sqrt
{2/3(v_0(x))}$ on the entire real line. Now substituting (\ref{k02})
into (\ref{N2}) we obtain
\begin{equation}\label{N3}
N \cong \frac{1}{2\pi}\int^{+\infty}_{-\infty}k_0(x)dx =
\frac{1}{\sqrt{6}\pi}\int^{+\infty}_{-\infty} v_0^{1/2}(x)dx \, ,
\end{equation}
which agrees with the IST result (\ref{N1}).

\subsection{Soliton-amplitude distribution function}

We note that the formula (\ref{N3}) in fact represents a particular
case of a more general expression, which is obtained by retaining
the parameter $\lambda$ in the modulation integral $I_1$
(\ref{int1}) so that instead of $k_0(x)$ in (\ref{N3}) one
introduces
\begin{eqnarray}
k_0(x,\la) &=&  \sqrt {(2/3)( v_0(x)- \la)} \quad \hbox{for} \quad x \in
[x_1, x_2]\,,  \nonumber \\
\hbox{and} \quad k_0(x,\la) &=&  0 \qquad \qquad \qquad \qquad \ \
\hbox{for} \quad  x \notin [x_1, x_2] \, , \label{kla}
\end{eqnarray}
where $x_1 < x_2$ are the roots of the equation $v_0(x)= \lambda$.
Thus we arrive at the continuous family of quantities
characterised by the parameter $\la \in [0, A]$
\begin{equation}\label{nl}
F(\lambda) \cong \frac{1}{2\pi} \int
_{-\infty}^{+\infty}k_0(x,\la)dx =
\frac{1}{\sqrt{6}\pi}\int^{x_2}_{x_1} \sqrt{v_0(x) - \lambda}\ dx \,,
\end{equation}

\begin{figure}[ht]
\vspace{0.5cm} \centerline{\includegraphics[width=10.0cm]{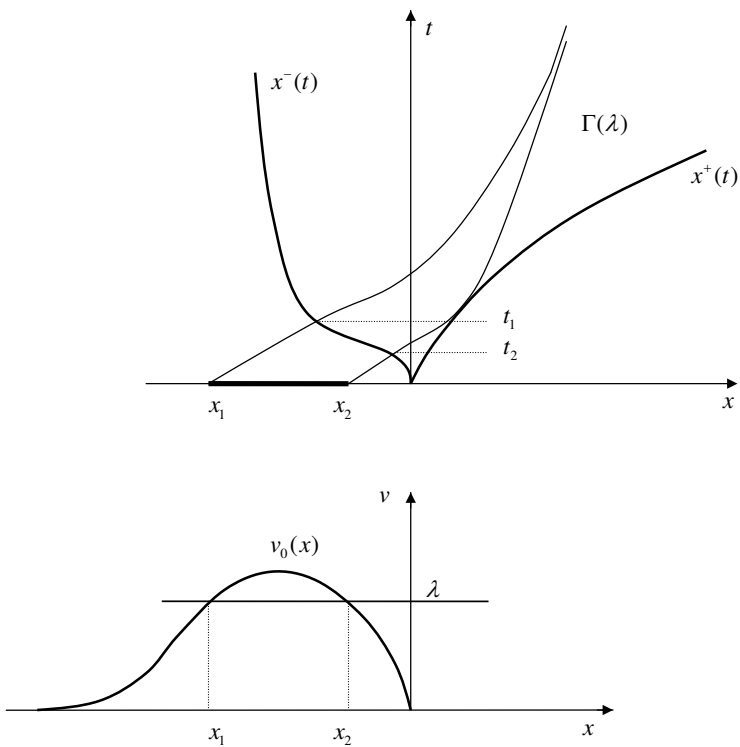}}
\vspace{0.3 true cm} \caption{$\la$-section of the initial profile
(below) and the behaviour of the corresponding confining
characteristics for part of the KdV modulation solution (above).
$\Gamma(\la)$ is the domain of influence of the interval $[x_1,
x_2]$} \label{fig1}
\end{figure}

To clarify the meaning of the quantity $F(\la)$ in the framework of
modulation theory we introduce the notion of a $\la$-section of the
initial profile, which is simply a segment of the function $v_0(x)$
for which $v_0(x) \ge \la$ (see Fig.~\ref{fig1}), and consider the
integral
\begin{equation}\label{Ft}
F(\la, t) = \frac{1}{2\pi}\int_{x_1 + \la t}^{x_2 + \la t} \sqrt
{(2/3)( \overline v(x,t)- \la)}dx
\end{equation}
for $t<t_2$, where $t_2$ is the time at which the characteristic
$x=x_2+\la t$ intersects the trailing edge $x^-(t)$ of the undular
bore $x^-(t_2)=x_2+\la t_2$ (see Fig.~\ref{fig1}). Note that this
$t_2$ plays for the $\la$-section of the (evolving) profile $v(x,t)$
a role similar to the breaking time, as the nonlinear
oscillations emerge in the vicinity of the point $(x^-(t_2), t_2)$
and the system (\ref{red1}) ceases to be valid. However, unlike at
the breaking point, there is no gradient catastrophe at $(x^-(t_2),
t_2)$.

We now see that for $t<t_2$ one has $\partial F/ \partial t =0$ and thus
$F(\la, t) = F(\la, 0) = F(\la)$. Indeed the integrand in (\ref{Ft}) is
nothing but the exact solution (\ref{ku}) of the modulation system
(\ref{red1}). But this system conserves the integral
$\int_{-\infty}^{+\infty} k(\overline v(x,t)) dx$, so that we have
\begin{equation}\label{conserv2}
\frac{1}{2\pi}\int_{-\infty}^{+\infty} k(\overline v(x,t)) dx =
\frac{1}{2\pi}\int_{x_1 + \la t}^{x_2 + \la t}k(\overline v(x,t)) dx
= \frac{1}{2\pi}\int_{x_1 }^{x_2 } k( v_0(x)) dx = F(\la)\,.
\end{equation}
It follows from (\ref{conserv2}) that $F(\la)$ is the number of linear
modulated waves (i.e.\ the number of wave crests) ``contained'' in
the $\la$-section of the initial profile. All these waves are
``released'' into the undular bore during the time interval $(t_2,
t_1)$ (see Fig.~\ref{fig1}) and eventually transform into solitons
as $t \to \infty$.

Now we establish a correspondence between the $\la$-section of the
initial profile and a certain part of the solitary wavetrain as $t
\to \infty$. This correspondence follows from the detailed
consideration of the behaviour of the characteristics for the
modulation solution made in Gurevich, Krylov \& Mazur \cite{gur89}
where it was shown that all the characteristics of the modulation
system issuing from the points of the interval $[x_1, x_2]$ are
confined to the area $\Gamma(\lambda)$ of $(x, t)$-plane enclosed by
the leading edge $x^+(t)$ and the characteristic emerging from $x_1$
(see Fig.~\ref{fig1}).  On the other hand, it was shown that for
sufficiently large $t$ the characteristics emerging from the points
$x \notin [x_1, x_2]$ lie entirely outside $\Gamma(\la)$. Therefore
all wave crests generated at the trailing edge between the points
$x^-_2=x_2+\la t_2$ and $x^-_1=x_1+\la t_1$ remain within the
described region $\Gamma(\lambda)$ for all $t$ and eventually
transform into the portion of solitons having their amplitudes in
the interval $[a(\lambda); a(A)]$. In other words, $\Gamma
(\lambda)$ represents the {\it domain of influence} of the interval
$[x_1, x_2]$  and thus, formula (\ref{nl}) defines the number of
solitons in this region as $t \to \infty$.  Indeed when $\la =0$ it
coincides with formula (\ref{N3}) for the total number of solitons.

We still need to find the dependence $a(\la)$ to obtain the distribution
of solitons as a function of amplitude.  In this regard we take advantage
of the fact that due to the condition $v(x,0)=0$ for $x>0$, all solitons
propagate on a zero background, so substituting $\overline v =0$ into $I_2$
we obtain $\lambda = 3\widetilde k^2/2$.  Now using the connection
(\ref{ak}) between the inverse width of a KdV soliton and its
amplitude we obtain the relationship $a=2\lambda$. In particular we
immediately recover the IST formula $a_{max}=a(A)=2A$ (see
(\ref{amax})). Then differentiating $F(\la)$ with respect to $a$ we
obtain Karpman's formula (\ref{karp}) for the soliton amplitude
distribution function
\begin{equation}\label{karpm}
f(a)=\left|\frac{dF(a/2)}{da} \right|=\frac{1}{4\pi}
\frac{1}{\sqrt{6}}\int_{x_1}^{x_2} \frac{dx}{\sqrt{v_0(x)- a/2}}\, .
\end{equation}

Summarising, we have managed to reproduce the semi-classical IST
results pertaining to soliton dynamics using, technically, only the
linear dispersion relation and the characteristic velocity of the
KdV equation in the dispersionless limit. Of course, we took
advantage of our knowledge of the qualitative behaviour of the
characteristics of the modulation equations for the solutions considered
here.  Next assuming the same qualitative behaviour of modulations
{\it as a plausible assumption}, we extend the above reasoning to the
non-integrable SG model (\ref{gn}) with the aim of obtaining the
counterparts of the formulae (\ref{N3}) and (\ref{karpm}) for fully
nonlinear shallow water dynamics. Our analytic results will then be
compared with full numerical solutions of the SG system.

\section{Asymptotic description of a solitary wavetrain in the
Su-Gardner system}

\subsection{Conserved quantities in the SG solitary wavetrain}

We consider the SG system (\ref{gn}) with the initial depth profile
\begin{equation}\label{depth}
\eta(x,0)=\eta_0(x)>1\, : \ \ \eta_0(x) \to 1  \quad \text{as} \quad
|x| \to \infty
\end{equation}
and the velocity profile connected with (\ref{vel}) by the simple wave
relationship
\begin{equation}
\label{vel}
u(x,0)=u_{0}(x) = 2(\sqrt{\eta_0(x)} -1) \, ,
\end{equation}
so that $u_0(x) \to 0$ as $|x| \to \infty$.  Let $\max [\eta_0(x)]=1+A$
and let $w$ be the typical width of $\eta_0(x)$.  Similarly to the KdV case
we shall assume that $A^{1/2}w \gg 1$. Also, for convenience, we
assume that $\eta_0(x)=1$ for $x>0$, which will guarantee that
solitary waves propagate on the undisturbed background $\eta=1$.

\begin{figure}[ht]
\vspace{0.5cm} \centerline{\includegraphics[width=8.0cm]{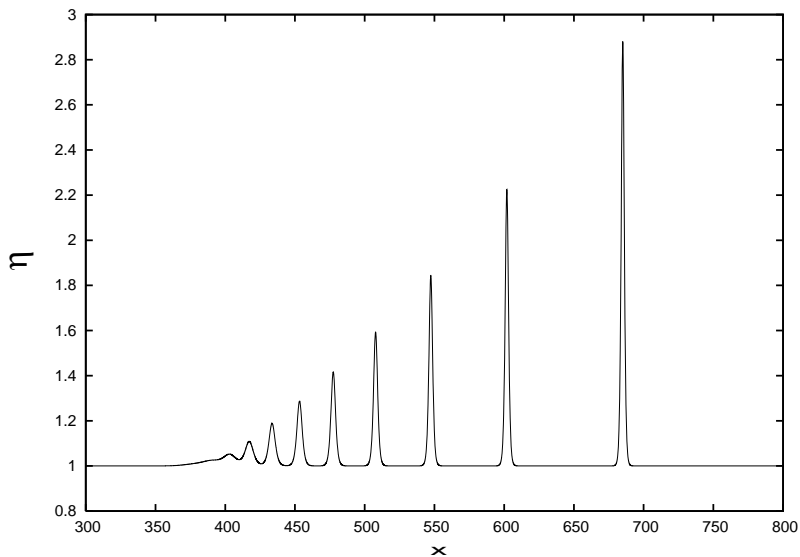}}
\vspace{0.3 true cm} \caption{Solitary wavetrain developing in the
SG system from the initial disturbance (\ref{init}) with $A=1$ and
$w=10$. The plot is $\eta(x,t)$ at $t=400$.}
\label{fig2}
\end{figure}

The system (\ref{gn}) conserves the four physical quantities
\begin{equation}\label{i1}
J_1=\int [\eta - 1] dx; \qquad \hbox{mass}
\end{equation}
\begin{equation}\label{i2}
J_2=\int [u+\frac{1}{6}\eta^2u_{xx}] dx; \qquad
\hbox{irrotationality}
\end{equation}
\begin{equation}\label{i3}
J_3= \int [\eta u] dx; \qquad \hbox{momentum}
\end{equation}
\begin{equation}\label{i4}
J_4=\int [\frac{1}{2}(\eta- 1)^2 + \frac{1}{2}\eta
u^2+\frac{1}{6}\eta^3 u_x^2]dx \qquad \hbox{energy}.
\end{equation}

We now consider the SG system with the specific initial conditions
\begin{equation}\label{init}
\eta_0(x) = 1 + A \ \hbox{sech}^{2}[ x/w], \qquad
u_0(x)=2(\sqrt{\eta_0(x)}-1)\,
\end{equation}
with the aim of verifying numerically some of our
main assumptions about the qualitative features of the asymptotic
behaviour as $t \to \infty$. A typical long-time outcome is shown in
Fig.~\ref{fig2}.  Comparison of the values of the four conserved quantities
(\ref{i1})--(\ref{i4}) computed for two initial profiles having the form
(\ref{init}) with $A=0.4$, $w=13$ and $A=2$, $w=10$ with their values
calculated for the solitary wavetrain at large $t$ shows that the
relative change, due to radiation, is $O(10^{-2})$, which is
within the accuracy of modulation theory, valid for
$\epsilon=(A^{1/2}w)^{-1} = 0.12$ and $0.07$ respectively. Therefore
we can assume that with good accuracy a positive large-scale
initial disturbance is completely transformed as $t \to \infty$
into a solitary wavetrain and we can neglect the contribution of
the radiation component in the modulation analysis.

\subsection{Modulation characteristic integrals}

We now apply the procedure for the derivation of the amplitude
distribution function, developed in Section 2 in the context of the
KdV equation, to the SG system. The key objects
are the modulation characteristic integrals, having in the
KdV case the form (\ref{int1}) and (\ref{int2}). The ordinary
differential equations for these integrals for the SG case (the
counterparts of the KdV case equations (\ref{ode1}) and (\ref{ode2}))
were derived in El, Grimshaw \& Smyth \cite{egs06}, so here we
present only the resulting formulae. It is important to mention that
the derivation in El, Grimshaw \& Smyth \cite{egs06} was performed
in the context of simple undular bores, so that it incorporated the
simple wave relationships between the depth and velocity jumps
across the bore. In the present study this condition is automatically
satisfied by choosing initial conditions in the form (\ref{depth})
and (\ref{vel}).

The key ingredient of the modulation characteristic integrals is the
linear dispersion relation for modulations, which in the SG case has
the form
\begin{equation}\label{lin}
a=0: \qquad \omega=\omega_0(\bar \eta, \bar u, k)=k\left(\bar u +
\frac{\bar \eta^{1/2}}{(1+\bar \eta^2k^2/3)^{1/2}}\right)\, .
\end{equation}
Here, as for $\overline v$ in (\ref{dr0}), $\overline \eta$ and
$\overline u$ denote the values averaged over the period of the
travelling wave.  Next we incorporate the simple wave relation
$\overline u (\overline \eta)=2(\sqrt{\overline \eta}-1)$ into
(\ref{lin}) to obtain the dispersion relation for a linear
dispersive wave propagating on a slowly varying simple wave
background
\begin{equation}\label{OM}
\Omega_0(\overline \eta, k)= \omega_0(\overline \eta, \overline u
(\overline \eta), k)=2k(\overline \eta^{1/2}-1)+ \frac{k\overline
\eta^{1/2}}{(1+\overline \eta^2k^2/3)^{1/2}} \, .
\end{equation}
Next, using reasoning identical to that described in Section 2
(see \cite{egs06,el05,ekt05} for additional details pertaining to bidirectional
systems), we obtain the characteristic integrals of the modulation system
defined by the ordinary differential equations
\begin{equation}\label{ode11}
I_1: \qquad a=0, \quad \frac{dk}{d\overline \eta}=\frac{\partial
\Omega_0 /
\partial \overline \eta}{V(\overline \eta) - \partial \Omega_0 /
\partial k} \quad \hbox{on}  \quad \frac{dx}{dt} = \frac{\partial
\Omega_0}{\partial k}\, ,
\end{equation}

\begin{equation}\label{ode21}
I_2: \qquad   k=0\, , \quad  \frac{d \widetilde k}{d\overline
\eta}=\frac{\partial \widetilde \Omega_0 /
\partial \overline \eta}{V(\overline \eta) - \partial  \widetilde \Omega_0
/ \partial \widetilde k} \quad \hbox{on}  \quad \frac{dx}{dt} =
\frac{\widetilde \Omega_0}{\widetilde k}\, .
\end{equation}
Here
\begin{equation}
\label{V}
V(\overline \eta)=\overline u(\overline \eta)+ \overline
\eta^{1/2}=3\overline \eta^{1/2}-2
\end{equation}
is the characteristic velocity of the right-propagating simple wave
of the ideal shallow water equations (i.e.\ the dispersionless limit of the SG
system) and
\begin{equation}
\label{conO}
\widetilde \Omega_0(\overline \eta, \widetilde k )=-i
\Omega_0(\overline \eta, i \widetilde k)= 2\widetilde k(\overline
\eta^{1/2}-1)- \frac{ \widetilde k\overline \eta^{1/2}}{(1 -
\overline \eta^2 \widetilde k^2/3)^{1/2}}\,
\end{equation}
is the SG ``solitary wave dispersion relation'', an analogue of the
KdV formula (\ref{conjomega}), $\widetilde k$ being the soliton
wavenumber.

Substituting (\ref{OM}) into (\ref{ode11}) and introducing
$\alpha=(1+k^2 \overline \eta^2/3)^{-1/2}$ as a new variable instead
of $k$ we obtain an ordinary differential equation in separated form
\begin{equation}\label{sep1}
\frac{d\overline \eta}{\overline
\eta}=\frac{2(1+\alpha+\alpha^2)}{\alpha(1+\alpha)(\alpha-4)}d\alpha\,
\, .
\end{equation}
Integrating (\ref{sep1}) we obtain
\begin{equation}\label{int11}
\frac{\overline\eta}{\lambda_1}
=\frac{1}{{\alpha}^{1/2}}\left(\frac{4-\alpha}{3}\right)^{21/10}
\left(\frac{1+\alpha}{2}\right )^{2/5}\, ,
\end{equation}
where $\la_1$ is an arbitrary constant of integration. Similarly,
substituting (\ref{conO}) into (\ref{ode21}) and then using
$\widetilde \alpha = (1 - {\widetilde k}^2 \overline
\eta^2/3)^{-1/2}$ instead of $\widetilde k$ we arrive at the same
separated ordinary differential equation (\ref{sep1}), but now for
for $\widetilde  \alpha (\overline \eta)$
\begin{equation}
\label{sep2}
\frac{d\overline \eta}{\overline \eta}=\frac{2(1+\widetilde \alpha+
\widetilde \alpha^2)}{\widetilde \alpha(1+ \widetilde
\alpha)(\widetilde \alpha-4)}d \widetilde \alpha
\end{equation}
with the integral
\begin{equation}\label{int21}
\frac{\overline\eta}{\lambda_2} =\frac{1}{{\widetilde
\alpha}^{1/2}}\left(\frac{4- \widetilde \alpha}{3}\right)^{21/10}
\left(\frac{1+ \widetilde \alpha}{2}\right )^{2/5}\, ,
\end{equation}
$\la_2$ being another constant of integration.

Now we recall that the constants $\la_1$ and $\la_2$ cannot be set
independently if the characteristic integrals (\ref{int11}) and
(\ref{int21}) are considered in the context of the same modulation
solution.  One can see from the zero amplitude integral $I_2$
(\ref{int11}) that $\overline \eta = \lambda_1$ implies $\alpha=1$
and, therefore, $k=0$. But $k=0$, in its turn, corresponds to the
solitary wave limit, so the equality $\overline \eta = \lambda_1$
must be consistent with the zero amplitude reduction of the
``soliton'' integral $I_2$ (\ref{int21}). To obtain this reduction
we calculate the solitary wave velocity using the conjugate
dispersion relation (\ref{conO})
\begin{equation}\label{cs1}
c=\frac{\widetilde \Omega_0(\overline \eta, \widetilde k
)}{\widetilde k} =2(\overline \eta^{1/2}-1)- \overline
\eta^{1/2}\widetilde \alpha \, .
\end{equation}
On the other hand, we have from (\ref{cs}), after replacing $\eta_0$
with $\overline \eta$ and $u_0$ with $\overline u = 2(\overline
\eta^{1/2}-1)$,
\begin{equation} \label{cs2}
c= 2(\overline \eta^{1/2}-1) + \sqrt{\overline \eta +a} \, .
\end{equation}
Comparing (\ref{cs1}) and (\ref{cs2}) we obtain the expression for the
solitary wave amplitude in terms of the variable $\widetilde
\alpha$, an analogue of the KdV formula (\ref{ak}),
\begin{equation}\label{as}
k=0: \qquad a=\overline \eta (\widetilde \alpha ^2 - 1)\, .
\end{equation}
Now one can see that the reduction as $k=0$, $a=0$ implies
$\widetilde \alpha =1$ which, by (\ref{int21}), immediately yields
$\overline \eta = \la_2$. Therefore, similar to the KdV case, we
have $\la_1=\la_2 \equiv \la$ and the relationships (\ref{int11}) and
(\ref{int21}) become a set of two consistent modulation
characteristic integrals parametrised by the same value $\la$ (cf.\
(\ref{int1}) and (\ref{int2}) for the KdV case).

\subsection{Total number of solitary waves}

As we described in Section 2.2 the total number of solitary waves
in the soliton train evolving out of the initial large-scale
disturbance (\ref{depth}) and (\ref{vel}) can be found from the
modulation formula
\begin{equation}\label{NSG}
N \cong \frac{1}{2\pi}\int _{-\infty}^{+\infty}k_0(x) dx \, ,
\end{equation}
where $k_0(x)$ is the initial distribution of the wavenumber
corresponding to the initial conditions (\ref{depth}) and (\ref{vel})
for the depth $\eta$ and velocity $u$. This distribution is found
from the group velocity characteristic integral of the
zero amplitude reduction of the modulation equations. In the case of
the SG system this is the integral $I_1$ given by formula
(\ref{int11}), in which one assumes $\lambda_1 = \la = 1$. The
latter equality follows from the requirement that the integral $I_1$
must correctly reproduce the mean value $\overline \eta =1$ for the
solitary wave background when one sets $k=0$ (i.e.\ $\alpha=1$) in
(\ref{int11}). Next one replaces $\overline \eta$ in $I_1$ with its
distribution $\eta_0(x)$ at $t=0$ to obtain an implicit expression
for $k_0(x)$ in terms of $\eta_0(x)$
\begin{equation}\label{int10}
\eta_0(x)
=\frac{1}{\sqrt{\alpha_0}}\left(\frac{4-\alpha_0}{3}\right)^{21/10}
\left(\frac{1+\alpha_0}{2}\right )^{2/5}\, ,
\end{equation}
where $\alpha_0(x)$ is connected with $k_0(x)$ via the relationship
\begin{equation}\label{a0}
\alpha_0=(1+(k_0\eta_0(x))^2/3)^{-1/2}\, .
\end{equation}

Thus expressing $k_0$ from (\ref{a0}) and substituting it into
(\ref{NSG}) we obtain
\begin{equation}
\label{NSG1}
N \cong \frac{1}{2\pi}\int _{-\infty}^{+\infty}
\frac{\sqrt{3(1-\alpha_0^2(x))}}{\eta_0(x)\alpha_0(x)}dx \, ,
\end{equation}
which, together with (\ref{int10}), completely defines the total
number of solitary waves forming in the initial value problem
(\ref{gn}), (\ref{depth}) and (\ref{vel}).

\begin{figure}[ht]
\vspace{0.5cm}
\centerline{\includegraphics[width=5.0cm]{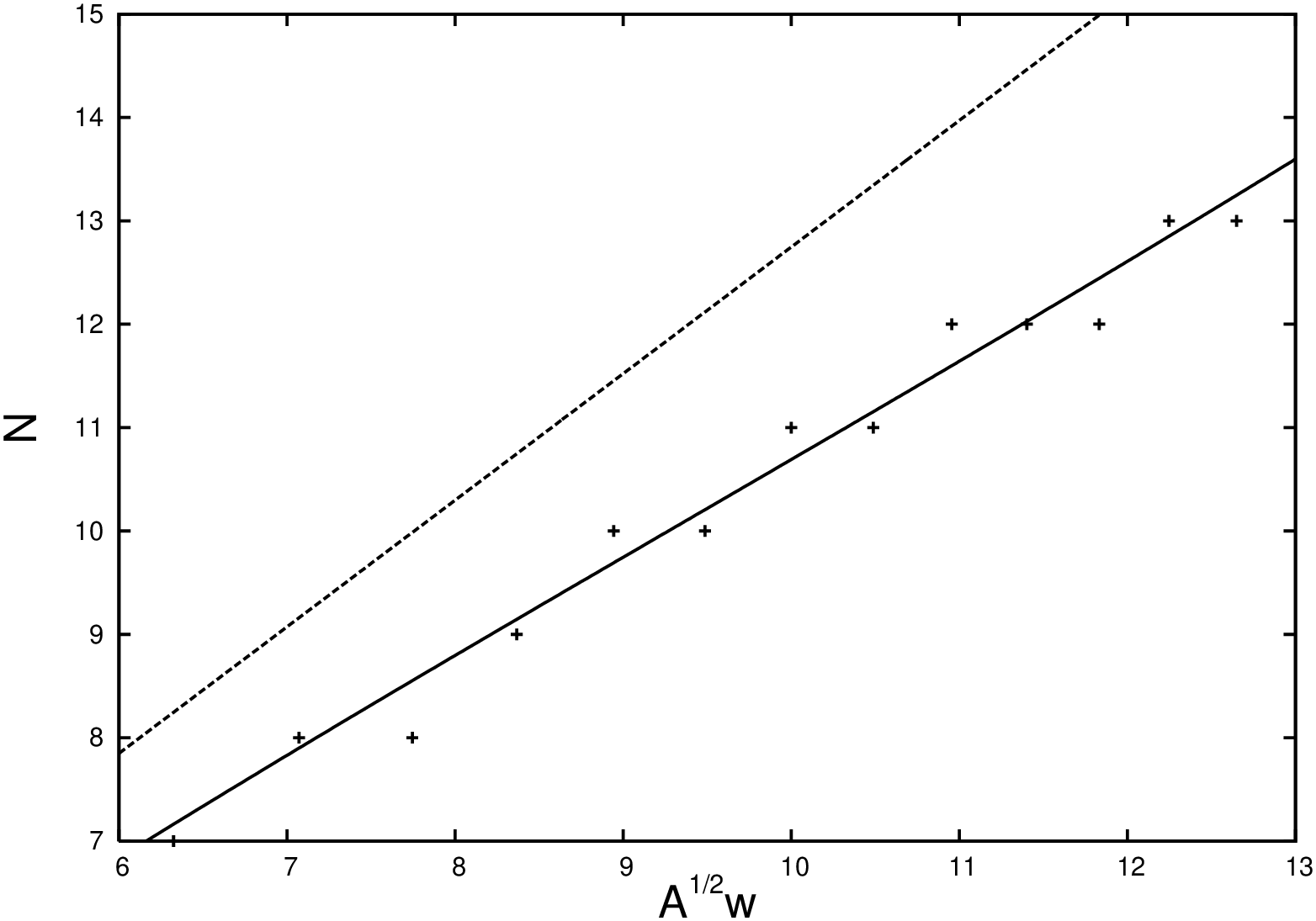} \qquad
\qquad \includegraphics[width=5.0cm]{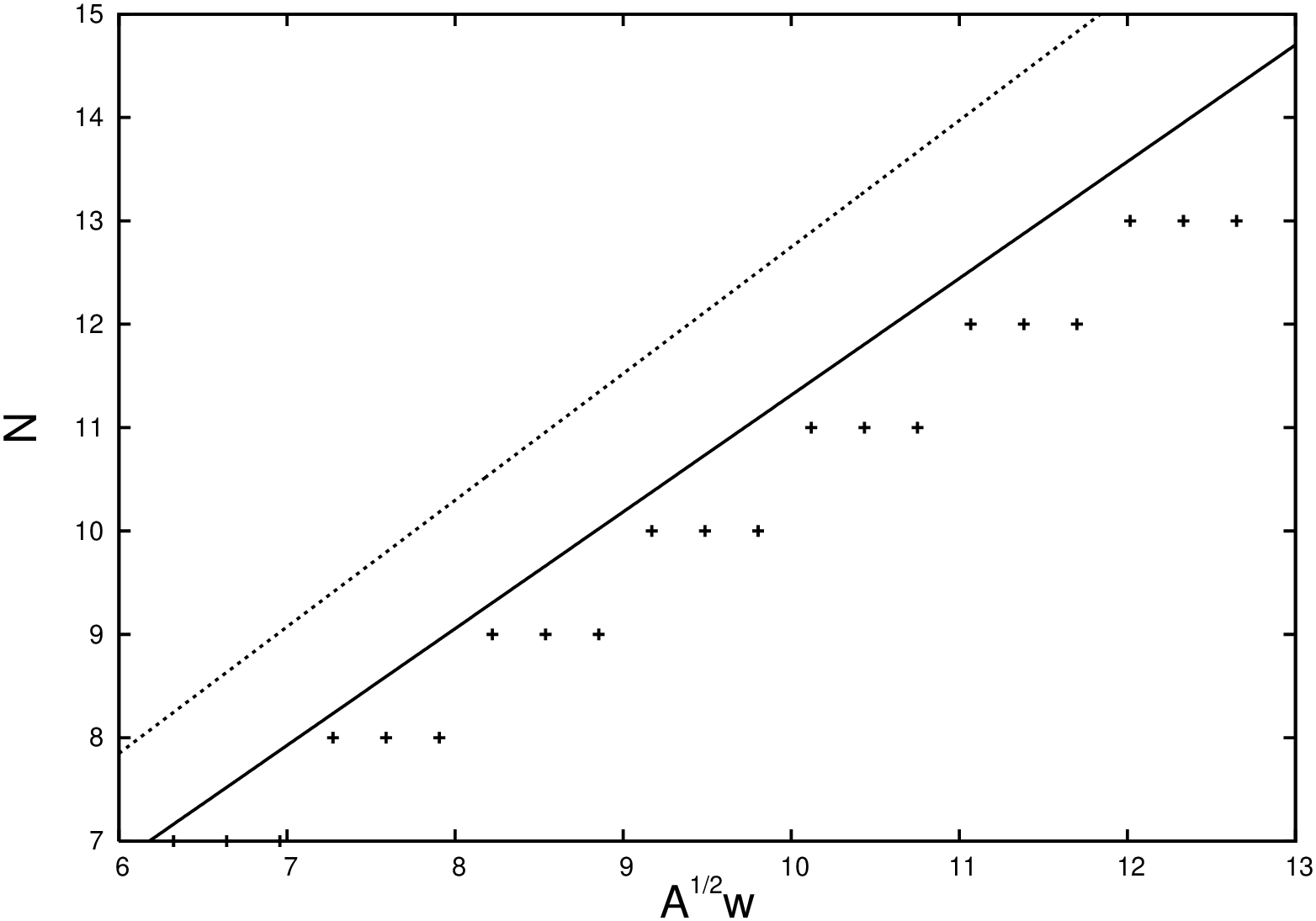}} \vspace{0.3 true
cm} \caption{Total number of solitary waves forming due to the decay of
the initial disturbance (\ref{init}). The initial parameters are
$w=10$ and $A$ varying (left);  $A=0.4$ and $w$ varying (right).
Solid line is the modulation formula (\ref{NSG1});
symbols are the numerical solution; the dashed line is the formula
(\ref{Nkdv}) corresponding to the KdV approximation (\ref{kdv}).}
\label{fig3}
\end{figure}

The first term of the expansion of (\ref{NSG1}) and (\ref{int10}) in
$\eta_0(x)-1 \ll 1$ (i.e.\  $1-\alpha_0 \ll 1$ by (\ref{int10})) yields
\begin{equation}
\label{Nkdv}
N \sim \sqrt{ \frac{3}{2}} \frac{1}{\pi}\int _{-\infty}^{+\infty}
\sqrt{\eta_0(x)-1} dx \, ,
\end{equation}
corresponding to the KdV result (\ref{N3}), taking into account that
the weakly nonlinear reduction of the SG system yields the KdV
equation in the form (\ref{kdv1}), so to obtain (\ref{N3}) one needs
firstly to set $\eta_0(x)-1=\zeta_0(x)$ in (\ref{Nkdv}) and then
apply the rescaling as described in Section 1.

To compare our analytical results with full numerical solutions of the
SG system we consider the dependence of $N$ on $A^{1/2}w$, which
is the parameter suggested by dimensional analysis (see also (\ref{N1})), for
initial data of the form (\ref{init}). The results are shown in Fig.~3. The
comparison shown in the left panel is for the solitary wavetrain
developing from the initial profile (\ref{init}), with the
width set to $w=10$ and the amplitude $A$ varying so that
$A^{1/2}w$ varies in the interval $[6, 13]$, while in the right panel
we fix the amplitude to $A=0.4$ and vary the width $w$ so that the
quantity $A^{1/2}w$ has the same range $[6, 13]$. One can see
that in both cases the modulation formula predicts the total number
of solitary waves very well. The over-prediction by just one in the
second comparison is clearly within the accuracy of the modulation
approach, which is $O(A^{1/2}w)^{-1})$.

\subsection{Amplitude distribution function}

Next, following the procedure of Section 2.3, we can ``upgrade''
formula (\ref{NSG1}) by using in (\ref{NSG}) the full expression
$k_0(x,\la)$, which is obtained from the modulation characteristic
integral $I_1$ (\ref{int11}) by retaining the parameter $\la$ and
replacing the mean depth $\overline \eta$ with its initial
distribution $\eta_0(x)$. The expected result is the SG analogue
of the integrated KdV soliton amplitude distribution function (\ref{nl}).

We thus assume that in (\ref{a0}) $\alpha_0=\alpha_0(x, \la)$
and then obtain from (\ref{int11}) an implicit expression
\begin{equation}\label{res2}
\frac{\eta_0(x)}{\la}
=\frac{1}{\sqrt{\alpha_0}}\left(\frac{4-\alpha_0}{3}\right)^{21/10}
\left(\frac{1+\alpha_0}{2}\right )^{2/5}\, .
\end{equation}
Then the integrated distribution function has the form
\begin{equation}\label{N4}
F(\la) \cong \frac{1}{2\pi} \int^{x_2}_{x_1}k_0(x, \la)dx \, ,
\end{equation}
where the function $k_0(x, \la)$ is expressed in terms of $\alpha_0(
x, \la)$ as
\begin{equation}\label{k0}
k_0(x, \la)=\frac{\sqrt{3(1-\alpha^2(x, \la))}}{\eta_0(x)\alpha(x,
\la)}\,
\end{equation}
(see(\ref{a0})) and $x_{1,2}(\la)$ are the roots of the equation
$k_0(x, \la)=0$. It immediately follows from (\ref{k0}) and (\ref{res2})
that these roots coincide with the roots of $\eta_0(x)=\la$. So the range
of $\la$ is $[1, 1+A]$, where $A=\max[\eta_0(x)]-1$.

\begin{figure}[ht]
\vspace{0.5cm} \centerline{\includegraphics[width=8.0cm]{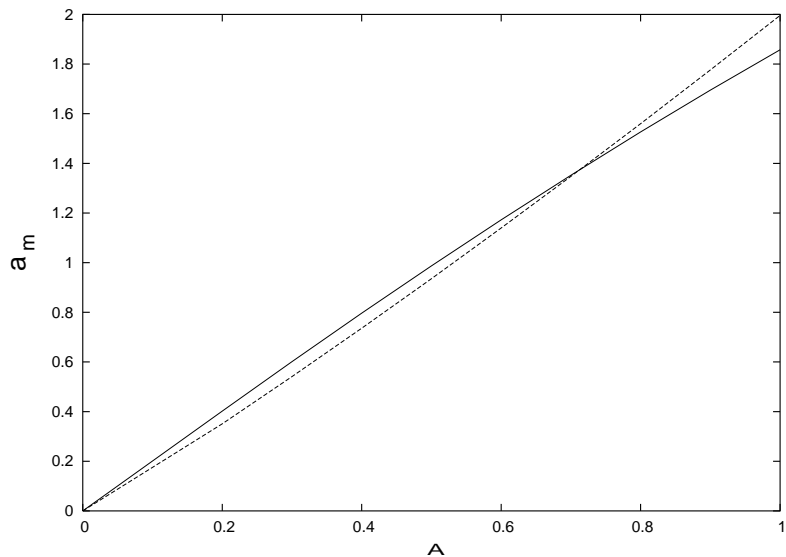}}
\vspace{0.3 true cm} \caption{Comparison of the lead solitary wave
amplitude $a_m$ as a function of the initial amplitude $A$.
The initial conditions are (\ref{init}) with $w=15$. Solid
line is the analytical solution (\ref{amSG}), the dashed line is the numerical
solution.}
\label{fig4}
\end{figure}

Next we establish the connection between the parameter $\la$ and
the lower bound of the amplitude range in the portion of solitary waves
corresponding to the ``$\la$-section'' of the initial profile $\eta_0(x)$.
This is found from the characteristic integral $I_2$ (\ref{int21}), in which
we set $\la_2=\la$ and $\overline \eta =1$ as here the solitary waves are on
a unit background (recall that $\eta_0(x)=1$ for $x>0$).
Then substituting $\widetilde \alpha= \sqrt{1+a}$ from (\ref{as})
we obtain the required relationship
\begin{equation}\label{la}
\la=(1+a)^{1/4}\left(\frac{3}{4-\sqrt{1+a}}\right)^{21/10}
\left(\frac{2}{1+\sqrt{1+a}}\right)^{2/5} \, .
\end{equation}

\begin{figure}[ht]
\vspace{0.5cm}
\centerline{\includegraphics[width=5.0cm]{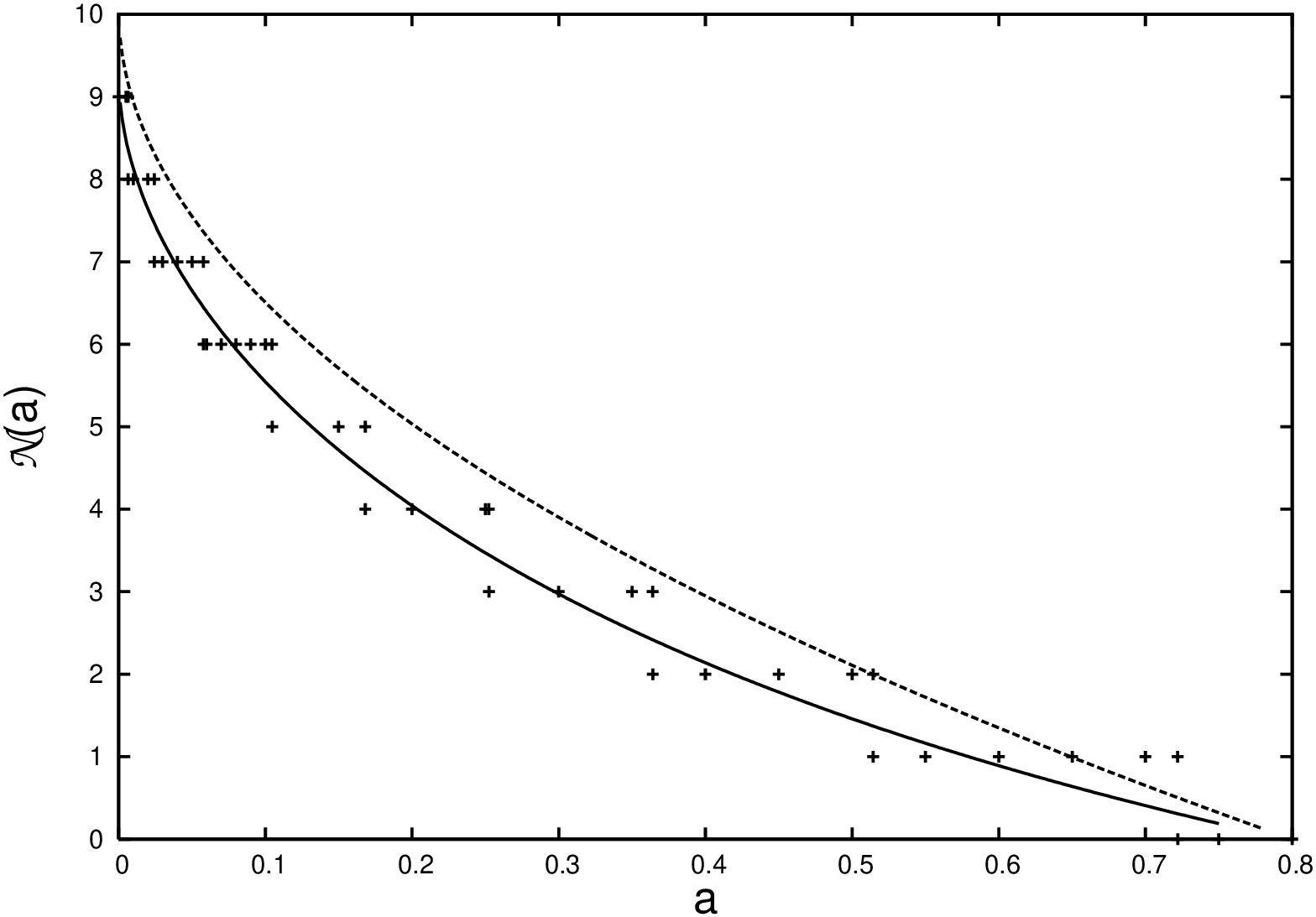}\qquad \qquad
\qquad
\includegraphics[width=5.0cm]{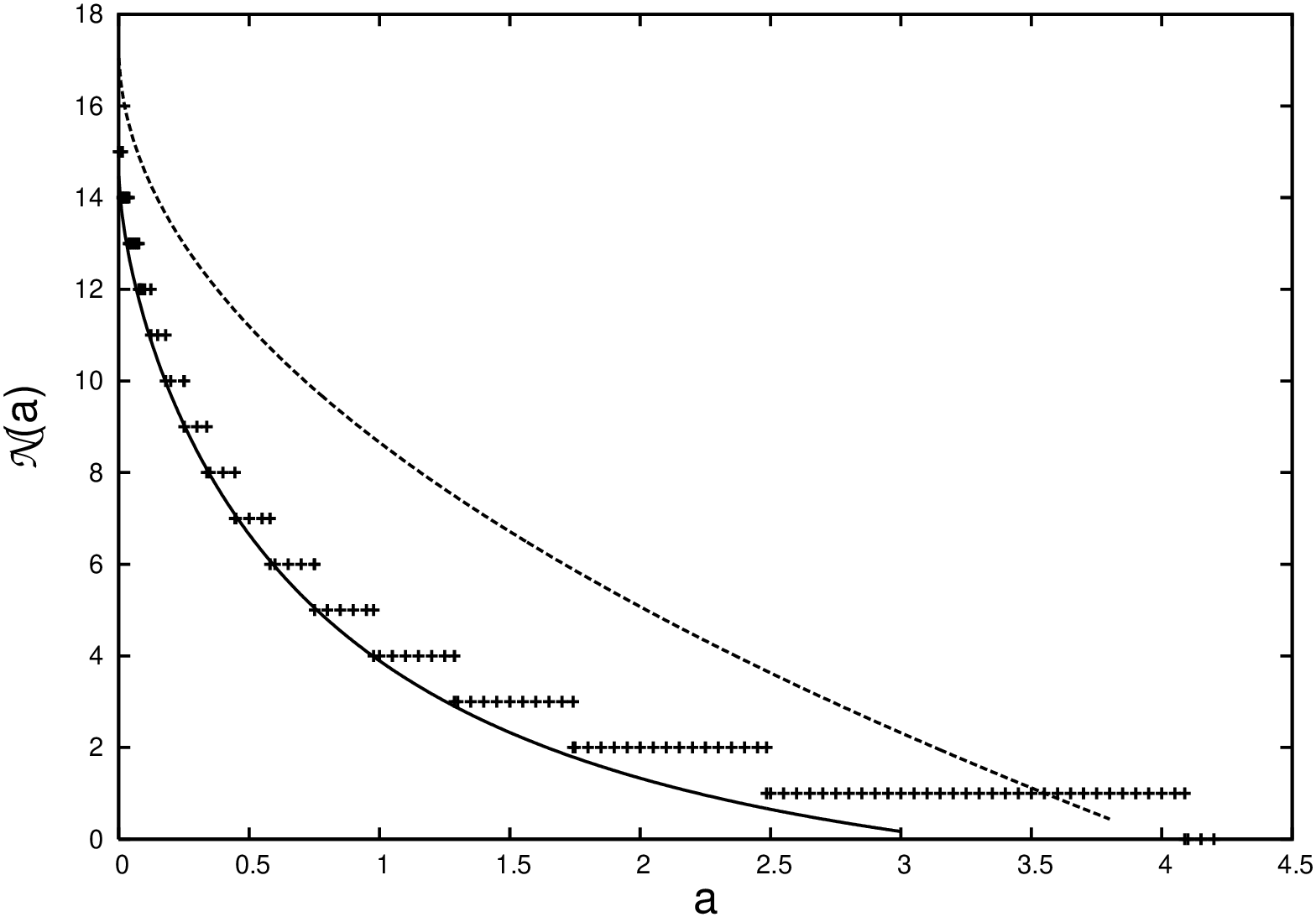}} \vspace{0.3 true cm}
\caption{Number $\mathcal{N}(a)$ of solitary waves with
amplitudes in the interval $[a, a_m]$ generated from the initial
conditions (\ref{init}) with $A=0.4$, $w=13$ (left) and $A=2$,
$w=10$ (right). Solid line is the modulation formula (\ref{N4}); symbols
are the numerical solution; dashed line is the integrated Karpman formula
(\ref{karp1}). }
\label{fig5}
\end{figure}

Since $\max\la=1+A$, the amplitude of the largest solitary wave
$a_{m}$ is found from the equation
\begin{equation}\label{amSG}
1+A=(1+a_{m})^{1/4}\left(\frac{3}{4-\sqrt{1+a_{m}}}\right)^{21/10}
\left(\frac{2}{1+\sqrt{1+a_m}}\right)^{2/5} \, .
\end{equation}
Expanding (\ref{amSG}) in $a_m$ we obtain to leading order $a_m =
2A$, as expected from weakly nonlinear KdV theory.  The relation
(\ref{amSG}) is shown in Fig.~\ref{fig4} along with the amplitude
curve obtained from direct numerical solutions of the SG system
(\ref{gn}) with the initial conditions (\ref{init}) with $w=15$.
One can see that the agreement is very good for solitary
wave amplitudes up to about $a=1.6$.  The departure of the analytical curve
from the numerical one for larger amplitudes will be discussed later. One
can also observe that the KdV lead soliton amplitude curve
$a_m=2A$ gives a very good approximation to the numerical
SG curve, in fact even better than the modulation SG formula (\ref{amSG}).
One should, however, bear in mind that formula (\ref{amSG}) for the
lead solitary wave amplitude should be used in conjunction with
expressions (\ref{sol}) and (\ref{cs}) defining the SG solitary
wave profile and velocity and, as such, then provides a consistent
approximation of the SG solution (and therefore of the full Euler
equations solution). At the same time the KdV formula $a_m=2A$
should be considered together with the familiar KdV soliton profile
and the speed-amplitude dependence (\ref{ca}) which are known not to
give very good approximations to the SG solitary wave for large
amplitudes.

\begin{figure}[ht]
\vspace{0.5cm}
\centerline{\includegraphics[width=8.0cm]{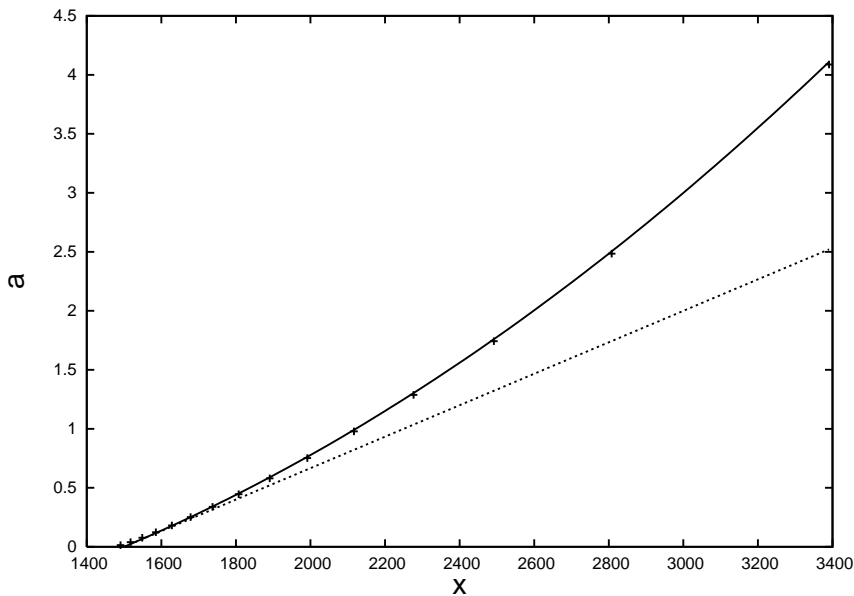}} \vspace{0.3
true cm} \caption{Amplitude profile of the SG solitary wavetrain at
$t=1500$. The parameters of the initial profile are $A=2$ and $w=10$.
Solid line is the formula (\ref{amp}); symbols are the numerical solution;
dotted line is the KdV soliton train amplitude profile (\ref{triangle})}
\label{fig6}
\end{figure}

Next putting (\ref{N4})--(\ref{la}) together we obtain the
formula for the number of solitary waves with amplitudes in the
interval $[a, a_m]$ as $\mathcal{N}(a)= F(\la(a))$. Indeed
$\mathcal{N}(0)= F(1) = N$, where $N$ is the total number (\ref{NSG1})
of solitary waves in the train.  Expanding (\ref{N4})--(\ref{la})
in $a$ for $a\ll 1$ we obtain to leading order
\begin{equation}
\label{karp1}
\mathcal{N}(a)  \sim  \sqrt{ \frac{3}{2}} \frac{1}{\pi}\int
_{x_1}^{x_2} \sqrt{(\eta_0(x)-1) - a/2} \ dx,
\end{equation}
where $x_{1,2}(a)$ are the roots of the equation $\eta_0(x)=1+a/2$.
Equation (\ref{karp1}) is the integrated Karpman formula for the KdV
equation in the form (\ref{kdv}) (cf.\ (\ref{nl})). Comparisons of
the integrated amplitude distribution function $\mathcal{N}(a)$
with the numerically found number of solitons are shown in
Fig.~\ref{fig5} for two different initial profiles (one with
$A=0.4$, $w=13$ (left panel) and another one with $A=2$, $w=10$
(right panel)). One can see that the agreement in both cases is
quite good (again taking into account the accuracy of the
modulation approach itself). Moreover, one can see from the
comparison in the right panel that the modulation formula appears to
work reasonably well far beyond the range of formal applicability of
the GP formulation of the problem used here. Indeed, it was shown in
El, Grimshaw and Smyth \cite{egs06} that starting from some critical depth
jump $\Delta_{cr} \approx 1.4$ across the SG undular bore, the
oscillatory structure of the bore qualitatively changes due to
linear degeneration of the characteristic field and formation of a
rapidly varying, finite amplitude rear wavefront, as opposed to the
usual vanishing amplitude trailing wave packet assumed in the GP
formulation.  As an estimate we can assume $A/2= \Delta - 1$, the
value of an ``effective jump'' taken at the level of half the
maximum value of the depth disturbance.  Then for solitary waves
with amplitudes greater than $a_{cr} \approx 0.8$ one can expect
some discrepancy between the modulation predictions and the results
of direct numerical simulations, as indeed seen in Fig.\ \ref{fig4}.

The solitary wave density distribution function $f(a)$ is obtained by
differentiating (\ref{karp1})
\begin{equation}\label{fa}
f(a)=\frac{d \mathcal{N}}{da}\, ,
\end{equation}
so that $f(a) da$ gives the number of solitons with amplitudes
in the interval $[a, a+da]$.  If we let the number of solitary waves per
unit length in the solitary wavetrain be $\kappa$, then it follows
from the balance relationship $\kappa dx=f(a) da$ that
$\kappa=f(a)da/dx$.  Since the speed of an SG solitary wave
propagating against a background $\eta=1$ and $u=0$ is
$c=\sqrt{1+a}$, the amplitude profile in the solitary wavetrain
for $x, t \gg 1$ is found from the general formula $\sqrt{1+a}=x/t+
\mathcal{O}(t^{-1})$ as
\begin{equation}\label{amp}
a\cong(x/t)^2-1 \, .
\end{equation}
A comparison of formula (\ref{amp}) for the solitary wavetrain amplitude
profile with the numerical profile for $A=2$ and $w=10$ at $t=1500$ is
shown in Fig.~\ref{fig6}

\begin{figure}[ht]
\vspace{0.5cm}
\centerline{\includegraphics[width=5.0cm]{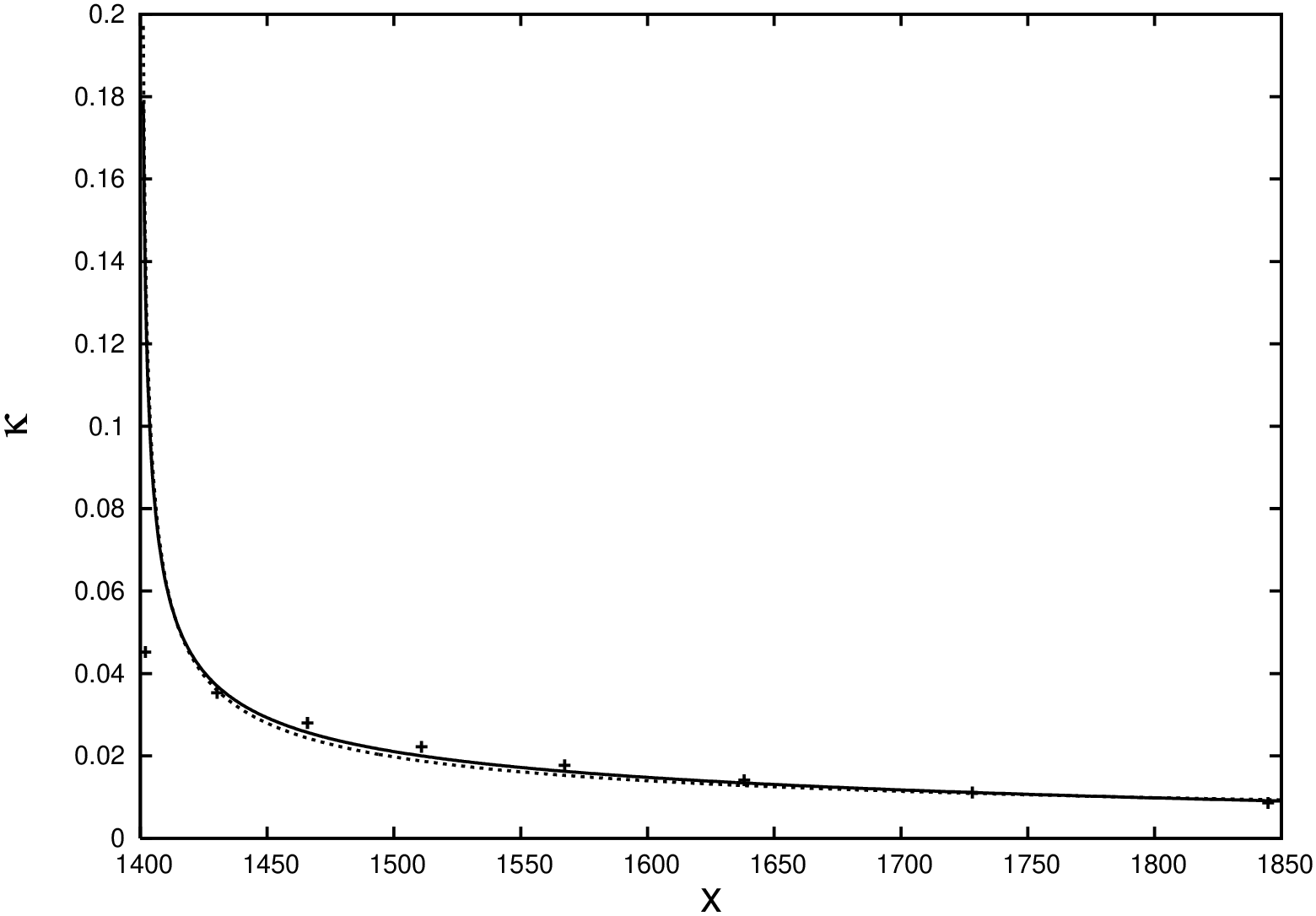} \qquad
\qquad \qquad \includegraphics[width=5.0cm]{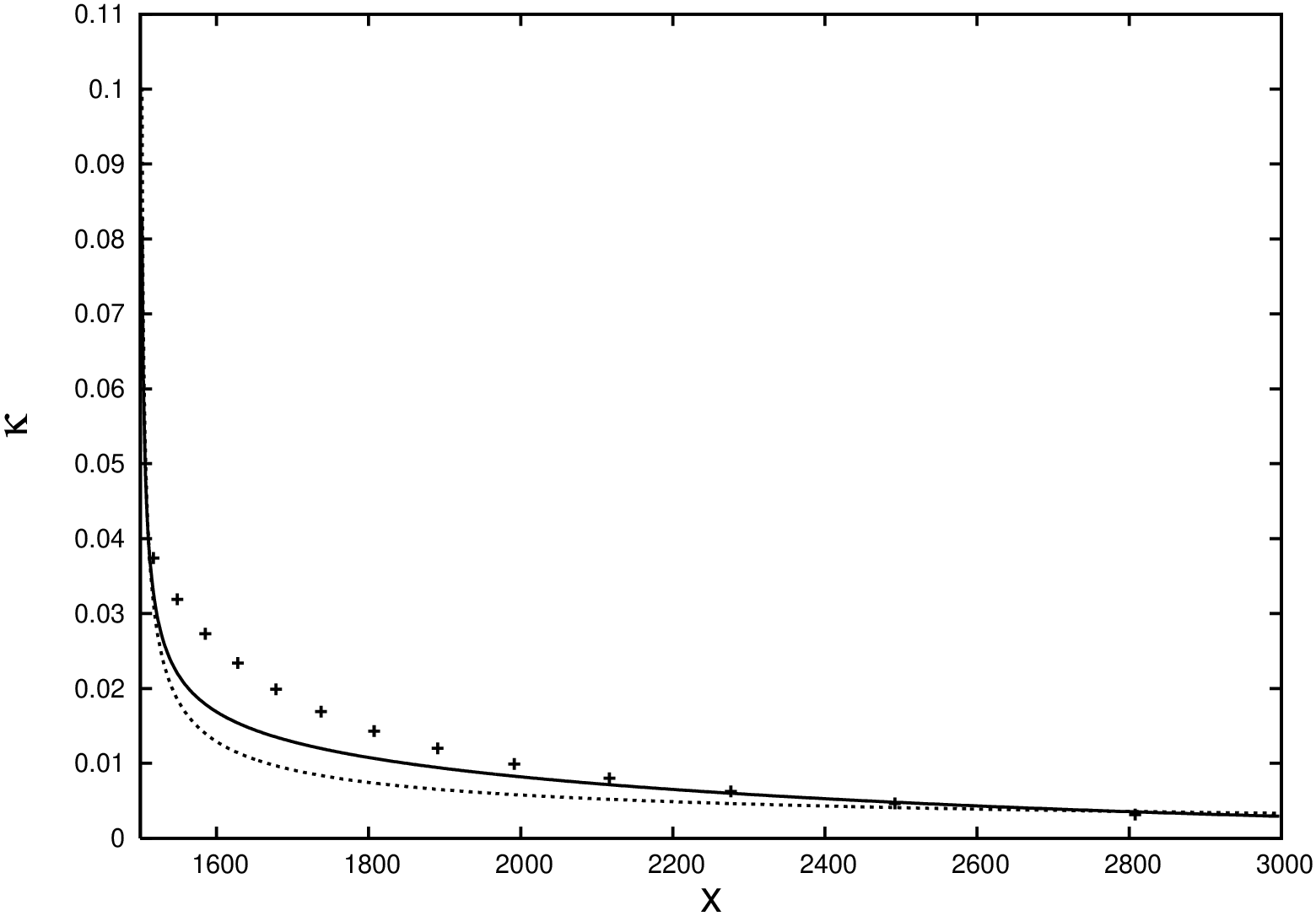}}
\vspace{0.3 true cm} \caption{Spatial density of solitary waves
$\kappa(x,t)$ in the solitary wavetrain formed out of the initial
profile (\ref{init}) with  $A=0.4$, $w=13$ (left, $t=1400$) and
$A=2$, $w=10$ (right, $t=1500$). Solid line is the SG formula
(\ref{kappa}); symbols are the numerical solution;
dotted line is the KdV formula(\ref{triangle}). }
\label{fig7}
\end{figure}

Next, for the number of solitons per unit length we obtain
\begin{equation}\label{kappa}
\kappa \cong \frac{2x}{t^2}f ( ({x}/{t})^2-1 )\, .
\end{equation}
We note that the weakly nonlinear counterparts of formulae
(\ref{amp}) and (\ref{kappa}) corresponding to the KdV equation in the
form (\ref{kdv}) are
\begin{equation}\label{triangle}
a\cong 2({x}/{t}-1  )\, , \qquad    \kappa \cong
\frac{2}{t}f_{KdV}(2(x/t-1))\, ,
\end{equation}
where $f_{KdV}(a)$ is given by the Karpman formula (\ref{karp}).
Comparisons of the curve (\ref{kappa}) with the numerically found
spatial density of solitary waves in the SG solitary wavetrain for
two different initial profiles are shown in Fig.~\ref{fig7}. As in
the previous comparisons, the agreement is excellent for moderate
($A=0.4$) and reasonably good for very large ($A=2$) amplitudes of
the initial disturbance (\ref{init}).

\section{Conclusions}

We have developed a general method for obtaining an asymptotic
description of solitary wavetrains for initial value problems for
weakly dispersive nonlinear systems which may not be integrable via
the IST. The method is based on the properties of the characteristics of
the associated modulation (Whitham) system describing an
intermediate undular bore stage of the evolution. We have
demonstrated the effectiveness of the developed approach by firstly
recovering the semi-classical IST results for the soliton amplitude
distribution function for the KdV equation and then by applying it
to the Su-Gardner (SG) system (\ref{gn}), the 1D version of the
Green-Naghdi equations, describing fully nonlinear shallow water
waves. The SG system is not integrable by the IST, but has enough
structure (a periodic travelling wave solution family and four
conservation laws) to be amenable to a nonlinear modulation analysis.
It also transpires that the SG system represents an important
mathematical model for understanding general properties of fully
nonlinear fluid flows beyond the strict shallow water limit. For
this reason we have studied its solutions for the full range of
amplitudes, although in the particular context of shallow water waves
the system (\ref{gn}) is unable to reproduce the effects of wave
overturning and becomes nonphysical for amplitudes greater than some
critical value.

The resulting asymptotic formulae for the solitary wave amplitude
distribution function, the total number of solitary waves and the
amplitude of the leading solitary wave in the solitary wavetrain
formed as the long time outcome of the decay of an initial
large scale, one hump depth disturbance have been compared with the
results of direct numerical simulations of the Su-Gardner system. Very
good agreement between the modulation solution and numerical
results have been demonstrated, even for the range of initial
amplitudes significantly exceeding that for the applicability of
modulation theory due to formation of a rapidly varying wavefront at
the trailing edge of the undular bore at an intermediate stage of its
evolution (see El, Grimshaw and Smyth \cite{egs06}).

The results obtained for fully nonlinear shallow water theory have
also been compared with their weakly nonlinear counterparts which are well
known from the semi-classical IST approach to the KdV equation,
where they were obtained as consequences of the famous
Bohr-Sommerfeld quantisation rule (see Whitham \cite{wh74} and Karpman
\cite{karp75}). Overall, one can conclude that KdV theory gives a very
good prediction for the lead solitary wave amplitude and the spatial
density of solitary waves in the fully nonlinear solitary wavetrain,
but consistently over-predicts the number of solitary waves in a
given amplitude interval.  Indeed this discrepancy grows with the initial
amplitude. One can also note that this comparison is consistent with the
modulation results for SG undular bores in the step resolution problem
studied in El, Grimshaw and Smyth \cite{egs06}, where it was shown that
the SG undular bore is noticeably narrower and contains less wavecrests
than its KdV counterpart for moderate to large values of the depth jump
across the bore. Also, due to the essentially different amplitude-speed
relationships for SG and KdV solitary waves (cf.\ (\ref{cs}) and (\ref{ca}),
the former being actually equivalent to that for the full Euler theory
\cite{ray}), the spatial amplitude profile of the SG solitary
wavetrain is parabolic, in contrast to the KdV classical triangle profile
(see Whitham \cite{wh74} and Fig.\ 6).

\vspace{0.5cm} \noindent {\bf Acknowledgements}

\noindent  The authors thank A.~Kamchatnov for useful discussions.

\end{document}